\documentclass[a4paper, 10pt, twoside, english]{book}
\newcommand{\Email}{Anna.Montagnini@univ-amu.fr}%
\newcommand{\Title}{Visual motion processing and human tracking behavior}%
\newcommand{\Acknowledgments}{%
The authors were supported by EC IP project FP7-269921, ``BrainScaleS'' and the French ANR-BSHS2-2013-006, ``Speed'.  They are grateful to Amarender Bogadhi, Laurent Madelain, Frederic Chavane and all the colleagues in the InVibe team for many enriching discussions. L.U.P wishes to thank Karl Friston and Rick Adams and the The Wellcome Trust Centre for Neuroimaging, University College London, for their essential contribution in the closed-loop delayed model. The chapter was written when G.S.M was an invited fellow from the CONICYT program, Chili. \\ Correspondence and requests for materials should be addressed to A.M. \\(email:\Email ).
}%

\usepackage{siunitx}%
\newcommand{\Hz}{\si{\hertz}}%
\newcommand{\ms}{\si{\milli\second}}%
\newcommand{\meter}{\si{\meter}}%
\usepackage{graphicx}
\usepackage[round]{natbib}
  {unsrtnat}%
\usepackage[breaklinks]{hyperref}
\author{Montagnini, Perrinet and Masson}\title{Human Motion Tracking}
\usepackage{fancyhdr}
\begin{document}
\frontmatter
 \addtocounter{chapter}{11} % set them to some other numbers than 0
\mainmatter
%\part{}
%\include{motiontracking}
%!TeX TS-program = Lualatex 
%!TeX encoding = UTF-8 Unicode 
%!TeX spellcheck = en-US
%!TEX root = main.tex
%\chapterauthor{\Authors}
\chapter{\Title}
%=========================================%
%______________%
%\section*{Abstract}
%\Abstract
\paragraph{BibTex entry}~~\\
\begin{verbatim}
@inbook{Montagnini15bicv,
    author = {Montagnini, Anna and Perrinet, Laurent U. and Masson, Guillaume S.},
    chapter = {12},
    citeulike-article-id = {13566752},
    editor = {Crist{\'{o}}bal, Gabriel and Perrinet, Laurent U. and Keil, Matthias S.},
    month = nov,
    publisher = {Wiley-VCH Verlag GmbH {\&} Co. KGaA},
    booktitle = {Biologically Inspired Computer Vision},
    keywords = {bicv-motion},
    posted-at = {2015-03-31 14:20:42},
    priority = {0},
    DOI = {10.1002/9783527680863.ch12},
    url = {http://onlinelibrary.wiley.com/doi/10.1002/9783527680863.ch12/summary},
    title = {Visual motion processing and human tracking behavior},
    year = {2015}
}
\end{verbatim}
\tableofcontents

%----------------------------------------------------%
\section{Introduction}
%----------------------------------------------------%

We are visual animals and vision is the primary source of information about the 3D layout of our environment or the occurrence of events around us. In nature, visual motion is abundant and generated by a large set of sources, such as the movement of another animal, be it predator or prey, or our own movements. Primates possess an highly performant system for motion processing. Such system is tightly linked to ocular tracking behaviors: a combination of smooth pursuit and saccadic eye movements is engaged to stabilize the retinal image of a selected moving object within the high-acuity region of our retinas, the fovea. During smooth phases of tracking, eye velocity nearly perfectly matches the target velocity, minimizing retinal slip. Fast saccadic eye movements interrupt these slow phases to correct position errors between the retinal location of the target image and the fovea. Thus, smooth pursuit and saccadic eye movements are respectively largely controlled by velocity and position information~\citep{Rashbass61}. Smooth pursuit eye movements (which are the focus of this chapter) are traditionally considered as the product of a \emph{reflexive} sensorimotor circuitry. Two important observations support this assumption: first, it is not possible to generate smooth pursuit at will across a stationary scene, and, second, it is not possible to fully suppress pursuit in a scene consisting solely of moving targets~\citep{Kowler11,Kowler14}. 

In non-human and human primates, visual motion information is extracted through a cascade of cortical areas spanning the dorsal stream of the cortical visual pathways (for reviews, see~\citep{MaunsellNewsome1987}). Direction and speed selective cells are found in abundance in two key areas lying at the junction between occipital and parietal lobes. Middle temporal (MT) and Medio-Superior Temporal (MST) areas can decode local motion information of a single object at multiple scales, isolate it from its visual background and reconstruct its trajectory~\citep{BradleyGoyal2008, BornBradley2005}. Interestingly, neuronal activities in areas MT and MST have been firmly related to the initiation and maintenance of tracking eye movements~\citep{NewsomeWurtz1988}. Dozens of experimental studies have shown that local direction and speed information for pursuit are encoded by MT populations~\citep{LisbergerMovshon1999} and transmitted to MST populations where non retinal information are integrated to form an internal model of object motion~\citep{Newsomeetal1988}. Such model is then forwarded to both frontal areas involved in pursuit control, such as frontal eye fields (FEF) and supplementary eye fields (SEF), as well as to the brainstem and cerebellar oculomotor system (see~\citep{Krauzlis2004} for a review).

What can we learn about visual motion processing by investigating smooth pursuit responses to moving objects? Since the pioneering work of Hassenstein and Reichardt~\citep{Hassenstein:56}, behavioural studies of tracking behavioural responses have been highly influential upon theoretical approaches to motion detection mechanisms (see Chapter \verb+017_tim_tiedemann+, this book), as illustrated by the fly vision literature (see~\citep{Borst:14, Borst-etal:10} for recent reviews) and its application in bio-inspired vision hardwares (e.g.~\citep{harrison-koch:00, kohler:09}). Because of their strong \emph{stimulus-driven} nature, tracking eye movements have been used similarly in primates to probe the properties of fundamental aspects of motion processing, from detection to pattern motion integration (see a series of recent reviews in~\citep{Masson2004,Kowler11, Lisberger2010}) and the coupling between the active pursuit of motion and motion perception~\citep{SperingMontagnini2011}. One particular interest of tracking responses is that they are time-continuous, smooth, measurable movements that reflect the temporal dynamics of sensory processing in the changes of eye velocity~\citep{Masson2004, Masson12}. Here, we would like to focus on a particular aspect of motion processing in the context of sensorimotor transformation: uncertainty.  Uncertainty arises from both random processes, such as the noise reflecting unpredictable fluctuations on the velocity signal, as well as from non stochastic processes such as ambiguity when reconstructing global motion from local information. We will show that both sensory noise and ambiguities impact the sensorimotor transformation as seen from the variability of eye movements and their course towards a steady, optimal solution.

Understanding the effects of various sources of noise and ambiguities can change our views on the two faces of the sensorimotor transformation. First, we can better understand how visual motion information is encoded in neural populations and the relationships between these population activities and behaviors~\citep{Lisberger2010,Osborne2011}. Second, it may change our view about how the brain controls eye movements and open the door to new theoretical approaches based on inference rather than linear control systems. Albeit still within the framework of stimulus-driven motor control, such new point of view can help us elucidate how higher level, cognitive processes, such as prediction, can dynamically interact with sensory processing to produce a versatile, adaptive behavior by which we can catch a flying ball despite its complex and even sometimes partially occluded trajectory. 

This chapter is divided in four parts. First, we will examine how noise and ambiguity both affect the pursuit initiation. In particular, we will focus on the temporal dynamics of uncertainty processing and the estimate of the optimal solution for motion tracking. Second, we will summarize how non-sensory, predictive signals can help maintaining a good performance when sensory evidences become highly unreliable and, on the contrary, when the future sensory inputs become highly predictable. Herein, we will illustrate these aspects with behavioral results gathered in both human and non-human primates. Third, we will show that these different results on visually guided and predictive smooth pursuit dynamics can be reconciled within a single Bayesian framework. Last, we will propose a biologically-plausible architecture implementing a hierarchical inference network for a closed-loop, visuomotor control of tracking eye movements.

%-------------------------------------------------------------------------%
\section{Pursuit initiation: facing uncertainties}
%-------------------------------------------------------------------------%

The ultimate goal of pursuit eye movements is to reduce the retinal slip of image motion down to nearly zero such that fine details of the moving pattern can be analyzed by spatial vision mechanisms. Overall, the pursuit system acts as a negative feedback loop where the eye velocity matches target velocity (such that the \textit{pursuit gain} is close to 1) to cancel the image motion on the retina. However, because of the delays due to both sensory and motor processing, the initial rising phase of eye velocity, known as pursuit initiation, is \textit{open-loop}. This means that during this short period of time (less than about 100~\ms) no information about eye movements is available to the system and the eye velocity depends only on the properties of the target motion presented to the subject. This short temporal window is ideal to probe how visual motion information is processed and transformed into an eye movement command~\citep{Lisberger87,Lisberger2010}. It becomes possible to map the different phases of visual motion processing to the changes in initial eye velocity and therefore to dissect out the contribution of spatial and temporal mechanisms of direction and speed decoding~\citep{Masson12}. However, this picture becomes more complicated as soon as one considers more naturalistic and noisy conditions for motion tracking, whereby, for instance, the motion of a complex-shaped extended object has to be estimated, or when several objects move in different directions. We will not focus, here, on the last problem, which involves the complex and time-demanding computational task of object segmentation~\citep{Masson2004,Schutzetal10}. In the next subsections, we focus instead on the nature of the noise affecting pursuit initiation and the uncertainty related to limitations in processing spatially-localized motion information.  

\subsection{Where is the noise? Motion tracking precision and accuracy}
\label{sec:noise}

Human motion tracking is variable across repeated trials and the possibility to use tracking behavior (at least during the initiation phase) to characterize visual motion processing across time and across all kinds of physical properties of the moving stimuli relies strongly on the assumption that oculomotor noise does not override the details of visual motion processing.
In order to characterize the variability of macaques' pursuit responses to visual motion signals, \citet{Osborne2005} analyzed monkeys' pursuit eye movements during the initiation phase --- here, the first $125$~ms after pursuit onset. On the basis of a principal component analysis of pursuit covariance matrix, they concluded that pursuit variability was mostly due to sensory fluctuations in estimating target motion parameters such as onset time, direction and speed, accounting for around $92\%$ of the pursuit variability. In a follow-up study, they estimated the time course of the pursuit system's sensitivity to small changes in target direction, speed and onset time~\citep{Osborne2007}. This analysis was based on pursuit variability during the first $300$~ms after target motion onset. Discrimination thresholds (inverse of sensitivity) decreased rapidly during open-loop pursuit and, in the case of motion direction, followed a similar time course to the one obtained from the analysis of neuronal activity in area MT~\citep{Osborne2004}.

Noise inherent to the kinematic parameters of the moving target is not the only source of uncertainty for visual motion processing. A very well-known and puzzling finding in visual motion psychophysics is that the speed of low-contrast moving stimuli is most often underestimated as compared to high contrast stimuli moving with exactly the same motion properties (see chapter \verb+009_series+). In parallel with the perceptual misjudgment, previous studies have shown that tracking quality is systematically degraded (with longer onset latencies, lower acceleration at initiation and lower pursuit gain) with low contrast stimuli~\citep{Spering2005}. This reduction of motion estimation and tracking accuracy when decreasing the luminance-contrast of the moving stimulus has been interpreted as evidence in favor of the fact that, when visual information is corrupted, the motion tracking system relies more on internal models of motion, or motion \emph{Priors}~\citep{Weiss02}.

\subsection{Where is the target really going?}
\label{sec:aperture}

The external world is largely responsible for variability. Another major source of uncertainty when considering sensory processing is ambiguity: a single retinal image can correspond to many different physical arrangements of the objects in the environment. In the motion domain, a well-known example of such input ambiguity is called "the aperture problem". When seen through a small aperture, the motion of an elongated edge (i.e. a one-dimensional -1D- change in luminance, see Figure~\ref{fig:aperture_SP}a, middle panel) is highly ambiguous. The same local motion signal can be generated by a infinite number of physical translations of the edge. Hans Wallach (see~\citep{Wuerger96} for an English translation of the original publication in German) was the first psychologist to recognize this problem and to propose a solution for it. A spatial integration of motion information provided by  edges with different orientations can be used to recover the true velocity of the pattern. Moreover, two-dimensional (2D) features such as corners or line-endings (whereby luminance variations occur along two dimensions, see Figure~\ref{fig:aperture_SP}a, right panel, for an example) can also be extracted through the same, small aperture as their motion is no longer ambiguous. Again, 1D and 2D motion signals can be integrated to reconstruct the two-dimensional velocity vector of the moving pattern. After several decades of intensive research at both physiological and behavioral levels, it remains largely unclear what computational rules are used by the brain to solve the 2D motion integration problem (see~\citep{MassonIlg2010} for a collection of review articles).

Indeed, several computational rules for motion integration have been proposed over the last 40 years (see~\citep{BradleyGoyal2008} for a review). In a limited number of cases, a simple vector averaging of the velocity vectors corresponding to the different 1D edge motions can be sufficient. A more generic solution, the Intersection-of-Constraints (IOC) is a geometrical solution that can always recover the exact global velocity vector from at least two moving edges with different orientations~\citep{fennema-thompson79, Adelson1982}. However, the fact that perceived direction does not always correspond to the IOC solution (for instance, for very short stimulus duration~\citep{YoWilson1992} or when a single 1D motion signal is present~\citep{Lorenceau1993,GoreaLorenceau1991}) has supported the role of local 2D features in motion integration. 

Several feedforward computational models have been proposed to implement these different rules~\citep{Wilsonetal1992,LofflerOrbach1999}. All these feedforward models have the same architecture. Motion integration is seen as a two-stage computation. The first stage, corresponding to cortical area V1 in primates, extracts local motion informations through a set of oriented spatiotemporal filters. This correspond to the fact that most V1 neurons respond to the direction orthogonal to the orientation of an edge drifting across their receptive field~\citep{Albright1984}. The local motion analyzers feed a second, integrative stage where pattern motion direction is computed. This integrative stage is thought to correspond to the extra-striate cortical area MT in primates. MT neurons have large receptive fields, they are strongly direction selective and a large fraction of them can unambiguously signal the pattern motion direction, regardless of the orientation of its 1D components~\citep{Movshon85,Albright1984}. Different nonlinear combinations of local 1D motion signals can be used to extract either local 2D motion cues or global 2D motion velocity vectors. 
Another solution proposed by~\citet{Perrinet12pred} is to consider that local motion analyzers are modulated by motion coherency~\citep{Burgi00}. This theoretical model shows the emergence of similar 2D motion detectors. 
These two-stage frameworks can be integrated into more complex models where local motion information is diffused across some retinotopic maps.

\subsection{Human smooth pursuit as dynamic readout of the neural solution to the aperture problem}
\label{sec:aperture_SP}
\begin{figure}%[ht!]%[p!]
\centering{
\includegraphics[width=1.\columnwidth]{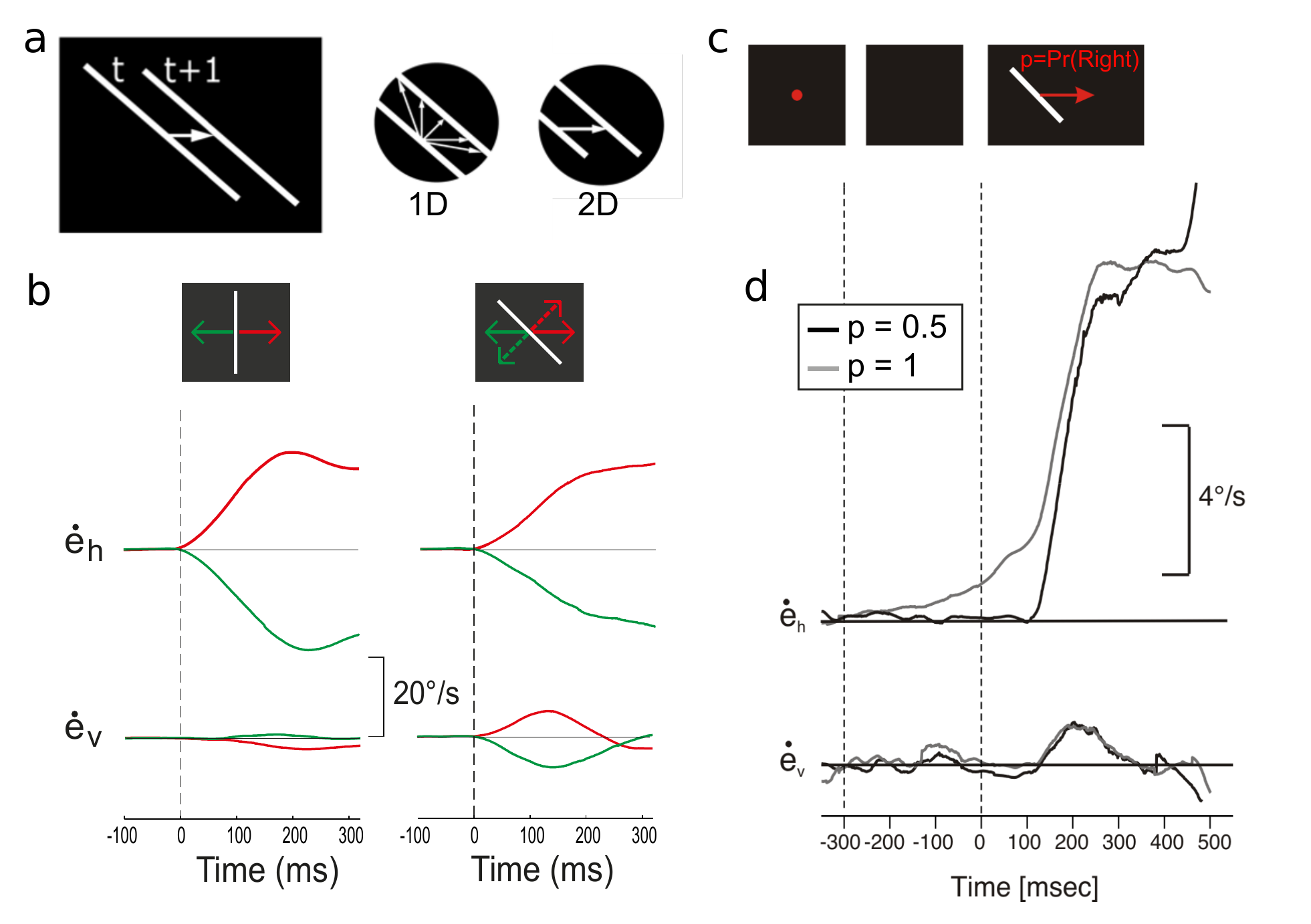}
}
\caption{Smooth pursuit's account for the dynamic solution of motion ambiguity and motion prediction: (a) A tilted bar translating horizontally in time (left panel) carries both ambiguous 1D motion cues (middle panel), and non-ambiguous 2D motion cues (rightmost panel). (b) Example of average horizontal ($\dot{e_h}$) and vertical ($\dot{e_v}$) smooth pursuit eye velocity while tracking a vertical (left) or a tilted bar (right) translating horizontally, either to the right (red curves) or to the left (green curves). Velocity curves are aligned on smooth pursuit onset. (c) Schematic description of a trial in the experiment on anticipatory smooth pursuit: after a fixation display, a fixed duration blank precedes the motion onset of a tilted line moving rightwards (with probability $p$) or leftward (with probability $1-p$). (d) Example of average horizontal ($\dot{e_h}$) and vertical ($\dot{e_v}$) smooth pursuit eye velocity in the anticipation experiment for two predictability conditions, $p=0.5$ (unpredictable, black curves) and $p=1$ (completely predictable, grey curves). 
\label{fig:aperture_SP}}% 
\end{figure}

Behavioral measures do not allow to capture the detailed temporal dynamics of the neuronal activity underlying motion estimate. However, smooth pursuit recordings do still carry the signature of the dynamic transition between the initial motion estimate dominated by the vector average of local 1D cues and the later estimate of global object motion. In other terms, human tracking data provides a continuous (delayed and low-pass filtered) dynamic readout of the neuronal solution to the aperture problem. Experiments in our and other groups~\citep{MassonStone2002, Wallace2005, Born2006} have consistently demonstrated, in both humans and monkeys, that tracking is transiently biased at initiation toward the direction orthogonal to the moving edge (or the vector average if multiple moving edges are present), when such direction does not coincide with the global motion direction. After some time (typically $200-300$~\ms) such bias is extinguished and the tracking direction converges to the object's global motion. In the example illustrated in Figure~\ref{fig:aperture_SP}, a tilted bar translates horizontally, thereby carrying locally ambiguous edge-related information (middle panel of part a). A transient non-zero vertical smooth pursuit velocity (lower right panel of Figure~\ref{fig:aperture_SP}b) reflects the initial aperture-induced bias, differently from the case where local and global motion are coherent (as for the pursuit of a horizontally moving vertical bar, see Figure~\ref{fig:aperture_SP}b, leftmost panels).     

The size of the transient directional tracking bias and the time needed for converging to the global motion solution depend on several properties of the visual moving stimulus~\citep{Wallace2005} including stimulus luminance contrast~\citep{Montagnini07,Bogadhi11}. In section~\ref{sec:Bayes_Kalman} we will see that this tracking dynamics is consistent with a simple Bayesian recurrent model (or equivalently a Kalman filter~\citep{Montagnini07}, which takes into account the uncertainty associated to the visual moving stimulus and combines it with prior knowledge about visual motion (see also chapter \verb+009_series+).

%------------------------------------------------------------------------%
\section{Predicting future and on-going target motion}
%------------------------------------------------------------------------%

Smooth pursuit eye movements can also rely on prediction of target movement to accurately follow the target despite a possible major disruption of the sensory evidence. Prediction allows also to compensate for processing delays,  an unavoidable problem of sensory-to-motor transformations. 

\subsection{Anticipatory smooth tracking}
\label{sec:anticipation}
The exploitation of statistical regularities in the sensory world and/or of cognitive information ahead of a sensory event is a common trait of adaptive and efficient cognitive systems that can, on the basis of such predictive information, anticipate choices and actions. It is a well-established fact that humans cannot generate smooth eye movements at will: for instance it is impossible to smoothly track an imaginary target with the eyes, except in the special condition in which an unseen target is self-moved through the smooth displacement of one's own finger~\citep{GauthierHofferer76}. In addition, smooth pursuit eye movements do necessarily lag unpredictable visual target motion by a (short) time delay. In spite of this, many years ago, already, it was known that, when tracking regular periodic motion, pursuit sensorimotor delay can be nulled and a perfect synchronicity between target and eye motion is possible (see~\citep{Kowler14,Barnes2008} for detailed reviews). Furthermore, when the direction of motion of a moving target is known in advance (for instance because motion properties are the same across many repeated experimental trials), anticipatory smooth eye movements are observed in advance of the target motion onset~\citep{Montagnini06}, as illustrated in the upper panel of Figure~\ref{fig:aperture_SP}d. Interestingly, also relatively complicated motion patterns, such as piecewise linear trajectories~\citep{BarnesSchmid2002}, or accelerated motion~\citep{Bennett2007} can be anticipated. Finally, probabilistic knowledge about target motion direction or speed~\citep{Montagnini2010}, and even subjectively experienced regularities extracted from the previous few trials~\citep{Kowleretal_1984} can modulate in a systematic way anticipatory smooth pursuit. 
Recently, several researchers have tested the role of higher-level cognitive cues for anticipatory smooth pursuit, leading to a rather diverse set of results. Although verbal or pictorial abstract cues indicating the direction of the upcoming motion seem to have a rather weak (though non inexistent) influence on anticipatory smooth pursuit, other cues are more easily and immediately interpreted and used for motion anticipation~\citep{Kowler14}. For instance, a barrier blocking one of two branches in a line-drawing illustrating an inverted-y-shaped tube, where the visual target was about to move~\citep{Kowler1989}, leads to robust anticipatory smooth tracking in the direction of the other, unblocked branch.

\subsection{If you don't see it, you can still predict (and track) it}
\label{sec:blank}
\begin{figure}%[ht!]%[p!]
\centering{
\includegraphics[width=1.\columnwidth]{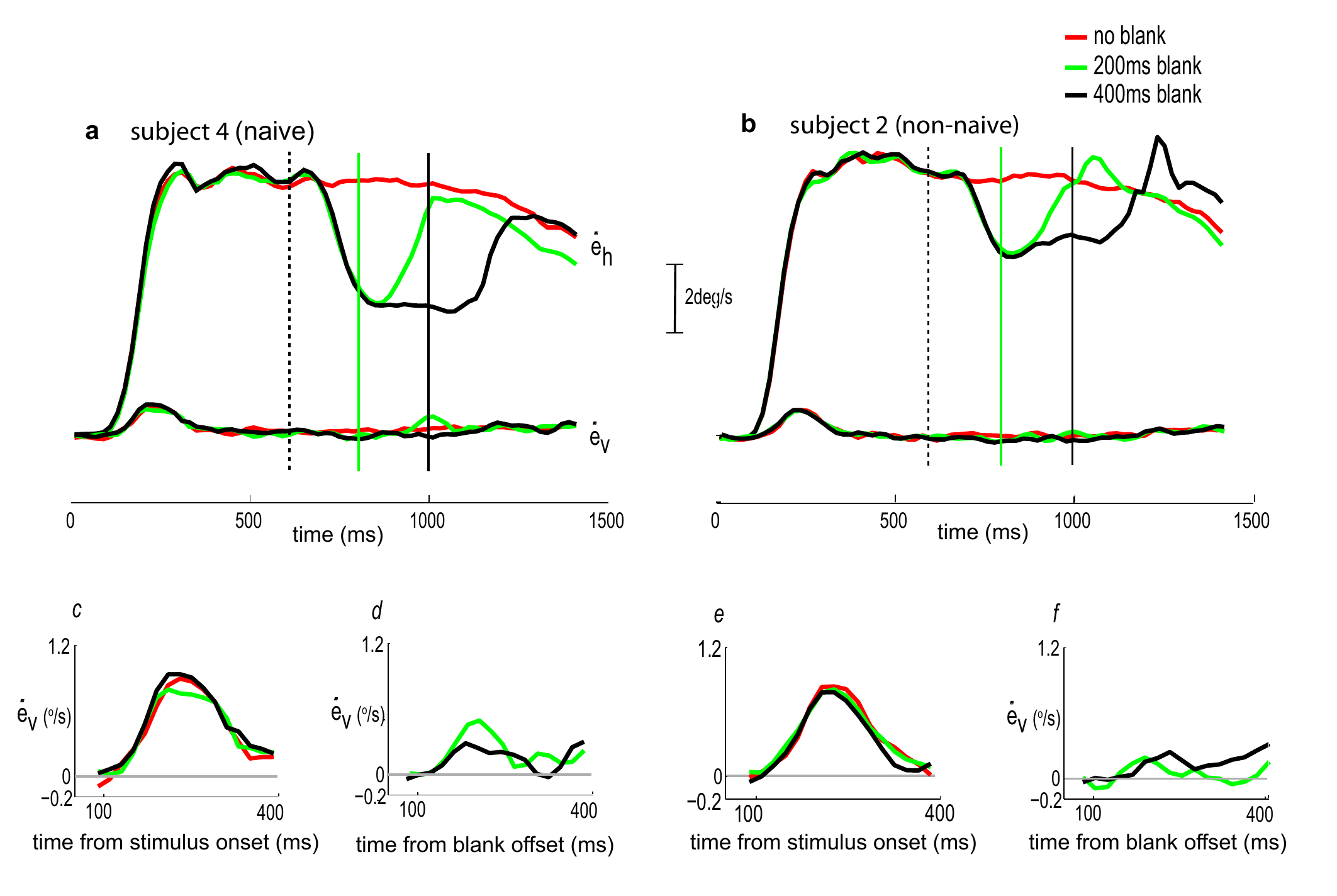}
}
\caption{Examples of human smooth pursuit traces (one different participant on each column, a naive one on the left and a non-naive one on the right side) during horizontal motion of a tilted bar which is transiently blanked during steady-state pursuit. (a) and (b): Average horizontal ($\dot{e_h}$) and vertical ($\dot{e_v}$) eye velocity. Different blanking conditions are depicted with different colors, as from figure legend. The vertical dashed line indicated the blank onset; vertical full colored lines indicate the end of the blanking epoch for each blanking duration represented. (c) and (e): Zoom on the aperture-induced bias of vertical eye velocity at target motion onset, for all blanking conditions. (d) and (f): Zoom on the aperture-induced bias of vertical eye velocity at target reappearance after blank (time is shifted so that 0 corresponds to blank offset), for all blanking conditions.
\label{fig:blank_traces}}% 
\end{figure}

While walking on a busy street downtown, it can occur that we track a moving car with our gaze and, despite it being hidden behind a truck occasionally driving in front of it, we can still closely follow its motion and have our gaze next to the car's position at its reappearance. In the lab, researchers have shown~\citep{Becker85} that during the transient disappearance of a moving target human subjects are capable to keep tracking the hidden motion with their gaze, although with a lower gain (see Figure~\ref{fig:blank_traces}, upper panels). During blanking, indeed, after an initial drop, eye velocity can be steadily maintained, typically at about 70\% of pre-blanking target velocity, although higher eye speed can be achieved with training~\citep{Madelain03}. In addition, when the blank duration is fixed, an anticipatory re-acceleration of the gaze rotation is observed ahead of target reappearance~\citep{BennettBarnes2003}. Extra-retinal, predictive information is clearly called into play to drive ocular tracking in the absence of a visual target. The true nature of the drive for such predictive eye velocity is still debated (see~\citep{Kowler11} for a review). Previous studies have proposed that it could either be a copy of the oculomotor command (an efference copy) serving as a positive feedback~\citep{Churchland2003,Madelain03} or a sample of visual motion being held in working memory~\citep{BennettBarnes2003}. In all cases a rather implausible "switch-like" mechanism was assumed, in order to account for the change of regime between the visual- and  prediction-driven tracking. 

While the phenomenology of human smooth pursuit during the transient absence of a visual target is well investigated (see for example~\citep{BennettBarnes2003,Orban2006,Orban2008}), less is known about its functional characterization, and about how the extra-retinal signals implicated in motion tracking without visual input interact with retinal signals across time. In particular, as motion estimation for tracking is affected by sensory noise and computational limitations (see sections~\ref{sec:noise} and~\ref{sec:aperture}), do we rely on extra-retinal predictive information in a way that depends on sensory uncertainty? In the past two decades, the literature on optimal cue combinations in multi-sensory integration has provided evidence for a weighted sum of different sources of information (such as visual and auditory~\citep{AlaisBurr2004}, or visual and haptic cues~\citep{ErnstBanks2002}, whereby each source is weighted according to its reliability (defined as inverse variance). Recently,~\citet{Khoei13} have shown that a predictive term on the smoothness of the trajectory is sufficient to account for the motion extrapolation part of the motion, however it lacked a mechanism to weight retinal and extra retinal information. In another recent study, we have tested the hypothesis that visual and predictive information for motion tracking are weighted according to their reliability and dynamically integrated to provide the observed oculomotor command~\citep{Bogadhi13}.

In order to do so, we have analyzed human smooth pursuit during the tracking of a horizontally-moving tilted bar that could be transiently hidden at different moments (blanking paradigm), either early, during pursuit initiation, or late, during steady-state tracking. By comparing the early and late blanking conditions we found two interesting results: first, the perturbation of pursuit velocity caused by the disappearance of the target was more dramatic for the late than the early blanking, both in terms of relative velocity drop and presence of an anticipatory acceleration before target reappearance. Second, a small, but significant, aperture-induced tracking bias (as described in section~\ref{sec:aperture_SP}) was observed at target reappearance after late but not early blanking. Interestingly, these two measures (the size of the tracking velocity reduction after blank onset and the size of the aperture bias after target disappearance) turned out to be significantly correlated across subjects for the late blanking conditions. 

We interpreted the ensemble of these results as an evidence in favor of dynamic optimal integration of visual and predictive information: at pursuit initiation, sensory variability is strong and predictive cues related to target- or gaze-motion dominate, leading to a relative reduction of both the effects of target blanking and of the aperture-bias. On the contrary, later on, sensory information becomes more reliable and the sudden disappearance of a visible moving target leads to a more dramatic disruption of motion tracking; coherently with this, the (re)estimation of motion is more strongly affected by the inherent ambiguity of the stimulus (a tilted bar). Finally, the observed correlation of these two quantities across different human observers indicates that the same computational mechanism (i.e. optimal dynamic integration of visual and predictive information) is scaled, at the individual level, in such a way that some people rely more strongly than others on predictive cues rather than on intrinsically noisy sensory evidence. Incidentally, in our sample of human volunteers, the expert subjects seemed to rely more on predictive information than the naive ones. 

In section~\ref{sec:model_hierarchical}, we will illustrate a model which is based on hierarchical Bayesian inference and is capable to qualitatively capture the human behavior in our blanking paradigm. A second important question is whether and how predictive information is affected by uncertainty as well. We will start to address this question in the next section.

%-----------------------------------------------------------------------------------------------------------------------------------%
\section{Dynamic integration of retinal and extra-retinal motion information: computational models}
%-----------------------------------------------------------------------------------------------------------------------------------%

\subsection{A Bayesian approach for open-loop motion tracking}
\label{sec:Bayes_Kalman}
\begin{figure}%[ht!]%[p!]
\centering{
\includegraphics[width=1.\columnwidth]{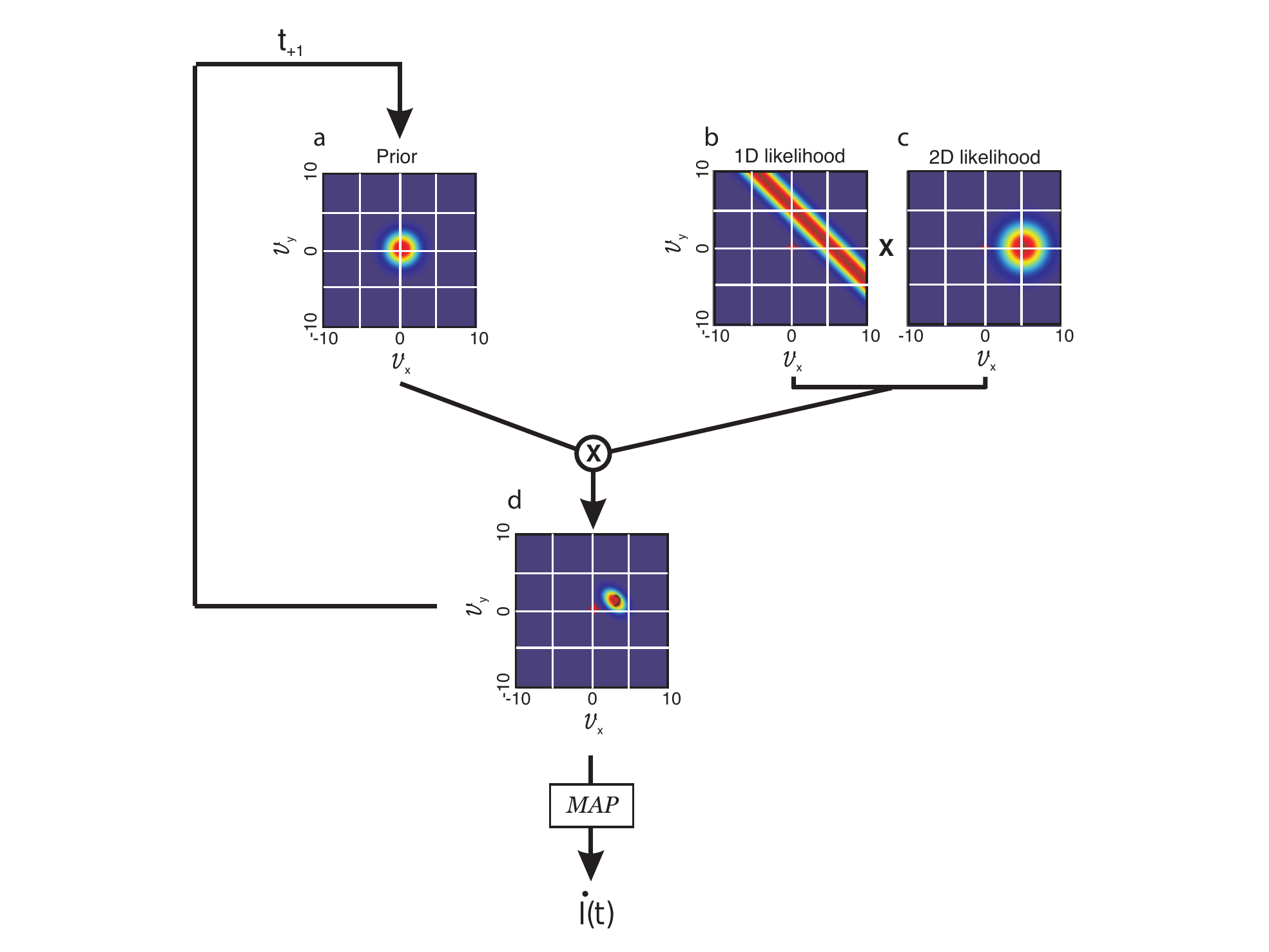}
}
\caption{A Bayesian recurrent module for the aperture problem and its dynamic solution. Top: ($a$) the Prior and the two independent 1D ($b$) and 2D ($c$) Likelihood functions (for a tilted line moving rightwards at $5$ deg/s) in the velocity space are multiplied to obtain the posterior velocity distribution (panel $d$). The inferred image motion is estimated as the velocity corresponding to the posterior maximum (MAP). Probability density functions are colour-coded, such that dark red corresponds to the highest probability and dark blue to the lowest one. 
\label{fig:retinal_recmodule}}% 
\end{figure}

Visual image noise and motion ambiguity, the two sources of uncertainty for motion estimate described in sections~\ref{sec:noise} and~\ref{sec:aperture} can be well integrated within a
Bayesian~\citep{Weiss02,Stocker06,Montagnini07,Perrinet07} or, equivalently, a Kalman-filtering  framework~\citep{Dimova2009,Orban2013}, whereby estimated motion is the solution of a dynamical statistical inference problem~\citep{Kalman60}. In these models, the information from different visual cues (such as local $1D$ and $2D$ motions) can be represented, as probability distributions, by their likelihood functions. Bayesian models also allow
the inclusion of prior constraints related to experience, expectancy bias and all possible sources of extra-sensory information. On the ground of a statistical predominance of static or slowly moving objects in nature, the most common
assumption used in models of motion perception is a preference for slow
speeds, typically referred to as low-speed prior (represented in panel $a$ of Figure~\ref{fig:retinal_recmodule}). The effects of priors are especially salient when signal uncertainty is high (see chapter \verb+009_series+). 

The sensory likelihood functions can be derived for
simple objects with the help of a few reasonable assumptions. For instance the motion cue associated with a non-ambiguous 2D feature would be approximated by a gaussian Likelihood centered on the true stimulus velocity and with a variance proportional to visual noise (e.g. inversely related to its visibility, see panel $c$ in Figure~\ref{fig:retinal_recmodule}). On the other hand, edge-related ambiguous information would be represented by an elongated velocity distribution parallel to the orientation of the moving edge, with an infinite variance along the edge direction reflecting the aperture ambiguity (panel $b$). Weiss and colleagues~\citep{Weiss02} have shown that, by combining a low-speed prior with an elongated velocity likelihood distribution parallel to the orientation of the moving line, it is possible to predict the aperture-induced bias (as illustrated in Figure~\ref{fig:retinal_recmodule}d). By introducing the independent contribution of the non-ambiguous 2D likelihood, as well as a recurrent optimal update of the Prior (with a feedback from the Posterior~\citep{Montagnini07}, see Figure~\ref{fig:retinal_recmodule}), and cascading this recurrent network with a realistic model of smooth pursuit generation~\citep{Bogadhi11}, we have managed to reproduce the complete dynamics of the solution of the aperture problem for motion tracking, as observed in human smooth pursuit traces.   

\subsection{Bayesian (or Kalman-filtering) approach for smooth pursuit: Hierarchical models}
\label{sec:model_hierarchical}
Beyond the inferential processing of visual uncertainties, that mostly affect smooth pursuit initiation, we have seen in section~\ref{sec:blank} that predictive cues can efficiently drive motion tracking when the visual information is deteriorated or missing. This flexible control of motion tracking has traditionally been modeled~\citep{Churchland2003,BennettBarnes2003,Madelain03,Orban2008} in terms of two independent modules, one processing visual motion and the other maintaining an internal memory of target motion. The weak point of these classical models is that they did not provide a biologically plausible mechanism for the interaction or alternation between the two types of control: a somewhat \emph{ad hoc} switch was usually assumed for this purpose.

Inference is very likely to occur at different spatial, temporal and neuro-functional scales. Sources of uncertainty can indeed affect different steps of the sensorimotor process in the brain. Recent work in our group~\citep{Bogadhi13} and other groups~\citep{Orban2013} has attempted to model several aspects of human motion tracking within a single framework, that of Bayesian inference, by postulating the existence of multiple inferential modules organized in a functional hierarchy and interacting according to the rules of optimal cue combination~\citep{Kalman60,Fetsch2012}. Here we outline an example of this modeling approach applied to the processing of visual ambiguous motion information under different conditions of target visibility (in the blanking experiment).

\begin{figure}%[ht!]%[p!]
\centering{
\includegraphics[width=1.\columnwidth]{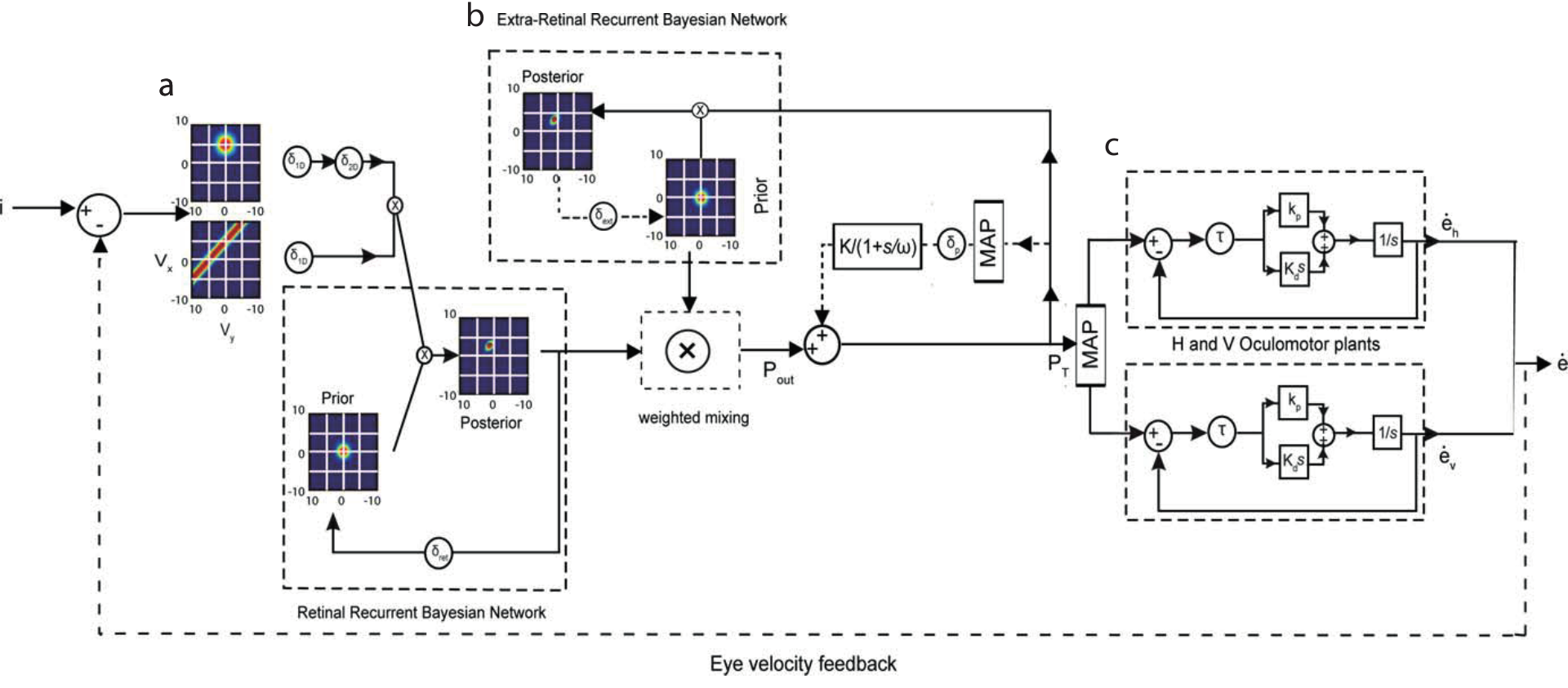}
}
\caption{Two-stages hierarchical Bayesian model for human smooth pursuit in the blanking experiment. The retinal recurrent loop ($a$) is the same as in figure~\ref{fig:retinal_recmodule}, with the additional inclusion of physiological delays. The posterior from the retinal recurrent loop and prior from the extra-retinal Bayesian network ($b$) are combined to form the post-sensory output ($P_{out}$). The maximum a posteriori of the probability ($P_T$) of target velocity in space
 serves as an input to both the positive feedback system as well as the oculomotor plants ($c$). The output of the oculomotor plant is
subtracted from the target velocity to form the image's retinal velocity (physical feedback loop shown as broken line). During the
transient blank when there is no target on the retina, the physical feedback loop is not functional so that the retinal recurrent block
does not decode any motion. The output of the positive feedback system (shown in broken line) is added to the post-sensory output
($P_{out}$) only when the physical feedback loop is functional. The probability distribution of target velocity in space ($P_{T}$) is provided as an
input to the extra-retinal recurrent Bayesian network where it is combined with a prior to obtain a posterior which is used to update
the prior.
\label{fig:two_stage_model}}% 
\end{figure}

In order to explain the data summarized in section~\ref{sec:blank} for the transient blanking of a translating tilted bar, we designed a two-modules hierarchical Bayesian recurrent model, illustrated in Figure~\ref{fig:two_stage_model}. The first module, the retinal recurrent network (panel $a$), implements the dynamic inferential process which is responsible for solving the ambiguity at pursuit initiation (see section~\ref{sec:aperture_SP}) and it only differs from the model scheme in Figure~\ref{fig:retinal_recmodule} for the introduction of processing delays estimated from the literature in monkey electrophysiology~\citep{Pack01}. The second module (panel $b$), the extra-retinal recurrent network, implements a dynamic short-term memory buffer for motion tracking. Crucially, the respective outputs of the retinal and extra-retinal recurrent modules are optimally combined in the Bayesian sense, so that their mean is weighted with their reliability (inverse variance) before the combination. By cascading the two Bayesian modules with a standard model~\citep{Goldreich92} for the transformation of the target velocity estimate into eye velocity (panel $c$) and adding some feedback connections (also standard in the models of smooth pursuit to mimic the closed-loop phase), the model is capable, with few free parameters, to simulate motion tracking curves that resemble qualitatively the ones observed for human subjects both during visual pursuit and during blanking. Note that one single crucial free parameter, representing a scaling factor for the rapidity with which the sensory variance increases during target blank, is responsible for the main effects described in section~\ref{sec:blank} on the tracking behavior during the blank and immediately after the end of it.

\Citet{Orban2013} have proposed an integrated Kalman filter model based on two filters, the first one extracting a motion estimate from noisy visual motion input, similar to a slightly simplified version of the previously described Bayesian retinal recurrent module. The second filter (referred to as \emph{predictive pathway}) provides a prediction for the upcoming target velocity on the basis of long term experience (i.e. from previous trials). Importantly, the implementation of a long term memory for a dynamic representation of target motion (always associated to its uncertainty) allows to reproduce the observed phenomenon of anticipatory tracking when target motion properties are repeated across trials (see section~\ref{sec:anticipation}. However in section~\ref{sec:open_questions} we will mention some results that challenge the current integrated models of hierarchical inference for motion tracking.

%-----------------------------------------------------------------------------------------------------------------------------------%
\subsection{A Bayesian approach for smooth pursuit: Dealing with delays}
%-----------------------------------------------------------------------------------------------------------------------------------%
\begin{figure}%[ht!]%[p!]
\centering{
\includegraphics[width=1.\columnwidth]{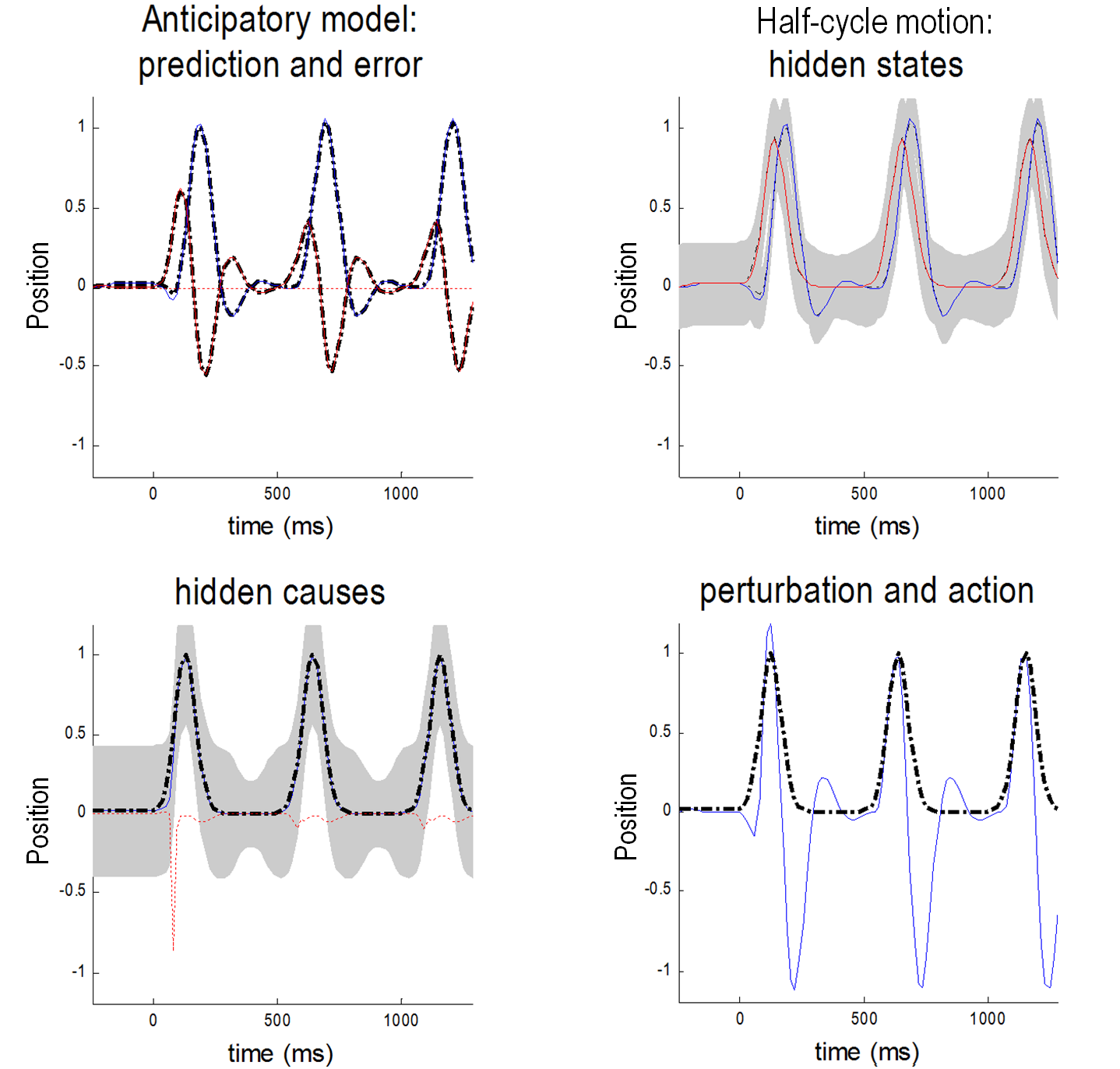}
}
\caption{
This figure reports the simulation of smooth pursuit when the target motion is hemi-sinusoidal, as would happen for a pendulum that would be stopped at each half cycle left of the vertical (broken black lines in the lower-right panel).
We report the horizontal excursions of oculomotor angle in retinal space (two upper panels) and the angular position of the target in an intrinsic frame of reference (visual space, lower panels).
The lower-right panel shows the true value of the displacement in visual space (broken black lines) and the action (blue line) which is responsible for
oculomotor displacements.
The upper left panel shows in retinal space the predicted sensory input (colored lines) and sensory prediction errors (dotted red lines) 
along with the true values (broken black lines). 
The latter is effectively the distance of the target from the centre of gaze
and reports the spatial lag of the target that is being followed (solid
red line). One can see clearly the initial displacement of the target
that is suppressed after a few hundred milliseconds. 
The sensory
predictions are based upon the conditional expectations of hidden
oculomotor (blue line) and target (red line) angular displacements shown
on the upper right. The grey regions correspond to 90\% Bayesian
confidence intervals and the broken lines show the true values of these
hidden states. 
The generative model used here has been
equipped with a second hierarchical level that contains hidden states,
modeling latent periodic behavior of the (hidden) causes of target
motion (states not shown here). 
The hidden cause of these displacements is shown with its
conditional expectation on the lower left. The true cause and action are
shown on the lower right. The action (blue line) is responsible for
oculomotor displacements and is driven by the proprioceptive prediction
errors.
\label{fig:PerrinetAdamsFriston14}
}% 
\end{figure}%

Recently, we considered optimal motor control and the particular problems caused by the inevitable delay between the emission of motor commands and their sensory consequences~\citep{PerrinetAdamsFriston14}. This is a generic problem that we illustrate within the context of oculomotor control where it is particularly 
insightful (see for instance~\citep{Nijhawan08} for a review). Although focusing on oculomotor control, the more general contribution of this work is to treat motor control as a pure inference problem. This allows us to use standard (Bayesian filtering) schemes to resolve the problem of sensorimotor delays --- by absorbing them into a generative (or forward) model. A generative model is a set of parameterized equations which describe our knowledge about the dynamics of the environment. Furthermore, this principled and generic solution has some degree of biological plausibility because the resulting active (Bayesian) filtering is formally identical to predictive coding, which has become an established metaphor for neuronal message passing in the brain (see~\citep{Bastos12} for instance). It uses oculomotor control as a vehicle to illustrate the basic idea using a series of generative models of eye movements --- that address increasingly complicated aspects of oculomotor control. In short, we offer a general solution to the problem of sensorimotor delays in motor control --- using established models of message passing in the brain --- and demonstrate the implications of this solution in the particular setting of oculomotor control.

Specifically, we considered delays in the visuo-oculomotor loop and their implications for active inference. Active inference uses a generalization of Kalman filtering to provide Bayes optimal estimates of hidden states and action (such that our model is a particular Hidden Markov Model) in generalized coordinates of motion. Representing hidden states in generalized coordinates provides a simple way of compensating for both sensory and oculomotor delays. The efficacy of this scheme is illustrated using neuronal simulations of pursuit initiation responses, with and without compensation. We then considered an extension of the generative model to simulate smooth pursuit eye movements --- in which the visuo-oculomotor system believes both the target and its centre of gaze are attracted to a (hidden) point moving in the visual field, similarly to what was proposed above in  section~\ref{sec:Bayes_Kalman}. Finally, the generative model is equipped with a hierarchical structure, so that it can recognize and remember unseen (occluded) trajectories and emit anticipatory responses (see section~\ref{sec:model_hierarchical}). 

We show in Figure~\ref{fig:PerrinetAdamsFriston14} the results of this model
for a two-layered hierarchical generative model.
%The first level of the generative model is similar to the above.
The hidden causes are informed by the dynamics of hidden
states at the second level: These hidden states model underlying
periodic dynamics using a simple periodic attractor that produces
sinusoidal fluctuations of arbitrary amplitude or phase and a frequency that
is determined by a second level hidden cause with a prior expectation of a frequency of $\eta$ (in \Hz).
It is somewhat similar to a control system model that attempts to
achieve zero-latency target tracking by fitting the trajectory to a
(known) periodic signal~\citep{Bahill83}. Our formulation
ensures a Bayes optimal estimate of periodic motion in terms of a posterior
belief about its frequency. In these simulations, we used a fixed Gaussian prior centered on the correct frequency with a period of $512~\ms$. This prior reproduces a typical experimental
setting in which the oscillatory nature of the trajectory is known, but
its amplitude and phase (onset) are unknown. Indeed, it has been shown
that anticipatory responses are cofounded when randomizing the
inter-cycle interval~\citep{Becker85}. In principle, we could
have considered many other forms of generative model, such as models
with prior beliefs about continuous acceleration~\citep{Bennett10}.
With this addition, the improvement in pursuit accuracy apparent
at the onset of the second cycle is consistent with what was observed empirically~\citep{Barnes00}. 

This is
because the model has an internal representation of latent causes of
target motion that can be called upon even when these causes are not
expressed explicitly in the target trajectory.
These simulations speak to a straightforward and neurobiologically plausible solution to the generic problem of integrating information from different sources with different temporal delays and the particular difficulties encountered when a system --- like the oculomotor system --- tries to control its environment with delayed signals. 
Neurobiologically, the
application of delay operators just means changing synaptic connection
strengths to take different mixtures of generalized sensations and their
prediction errors. 

%-----------------------------------------------------------------------------------------------------------------------------------%
\section{Reacting, inferring, predicting: a neural workspace}
%-----------------------------------------------------------------------------------------------------------------------------------%
We have proposed herein a hierarchical inference network that can both estimate the direction and speed of a moving object despite the inherent ambiguities present in the images and predict the target trajectory from accumulated retinal and extra-retinal evidence. What could be the biologically-plausible implementation of such a hierarchy? What are its advantages for a living organism? 

The fact that the pursuit system can be separated into two distinct blocks has been proposed by many others (see for recent reviews~\citep{Barnes2008,Lisberger2010, MassonIlg2010b,Fukushima13}. Such structure is rooted on the need to mix retinal and extra-retinal information to ensure stability of  pursuit, as originally proposed by~\citep{YasuiYoung:75}. Elaborations of this concept have formed the basis of a number of models based on a negative feedback control system~\citep{Robinson86,KrauzlisLisberger:94}. However, the fact that a simple efference copy feedback loop can not account for anticipatory responses during target blanking as well as for the role of expectation about future target trajectory~\citep{Barnes2008} or reinforcement learning during blanking~\citep{Madelain03} has called into question the validity of this simplistic approach (see~\citep{Fukushima13} for a recent review). This has led to more complex models where an internal model of target motion is reconstructed from an early sampling and storage of target velocity and an efference copy of the eye's velocity signal. Several memory components have been proposed to serve different aspects of cognitive control of smooth pursuit and anticipatory responses~\citep{BarnesCollins:11,Fukushima13}. The Hierarchical Bayesian model presented above (see section~\ref{sec:model_hierarchical}) follows the same structure with two main differences. First, the sensory processing itself is seen as a dynamical network whereas most of the models cited in this section have oversimplified target velocity representation. Second, we collapse all the different internal blocks and loops into a single inference loop representing the perceived target motion. 

We have proposed earlier that these two inference loops might be implemented by two large-scale brain networks~\citep{MassonIlg2010b}. The dynamical inference loop is based on the properties of primate visual areas V1 and MT where local and global motion signals have been clearly identified, respectively (see~\citep{MassonIlg2010} for a series of recent review articles). MT neurons solve the aperture problem with a slow temporal dynamics. When presented with a set of elongated, tilted bars, their initial preferred direction matches the motion direction orthogonal to the bar orientation. From there, that preferred direction gradually rotate toward the true, 2D translation of the bar so that after about $120$~\ms, MT population signals the correct pattern motion direction, independently of the bar orientation~\citep{Pack01}. Several models have proposed that such dynamics is due to reccurent interactions between V1 and MT cortical areas (e.g.~\citep{Tlapale10,Bayerl2004,Berzhanskaya07}) and we have shown that the dynamical Bayesian model presented above give a good description of such neuronal dynamics and its perceptual and behavioural counterparts~\citep{Montagnini07} (see also chapter \verb+010_keil+). Such recurrent network can be upscaled  to include other cortical visual areas involved in shape processing (e.g. areas V2, V3, V4) to further improve form-motion integration and select one target among several distractors or the visual background~\citep{Tlapale10}. The V1-MT loop exhibits however two fundamental properties with respect to the tracking of object motion. First, neuronal responses stop immediately when the retinal input disappears as during object blanking. Second, the loop remains largely immune to higher, cognitive inputs. For instance, the slow speed Prior used in our Bayesian model can hardly be changed by training in human observers~\citep{Montagnini06}.

\begin{figure}[ht!]%[p!]
\centering{
\includegraphics[width=1.\columnwidth]{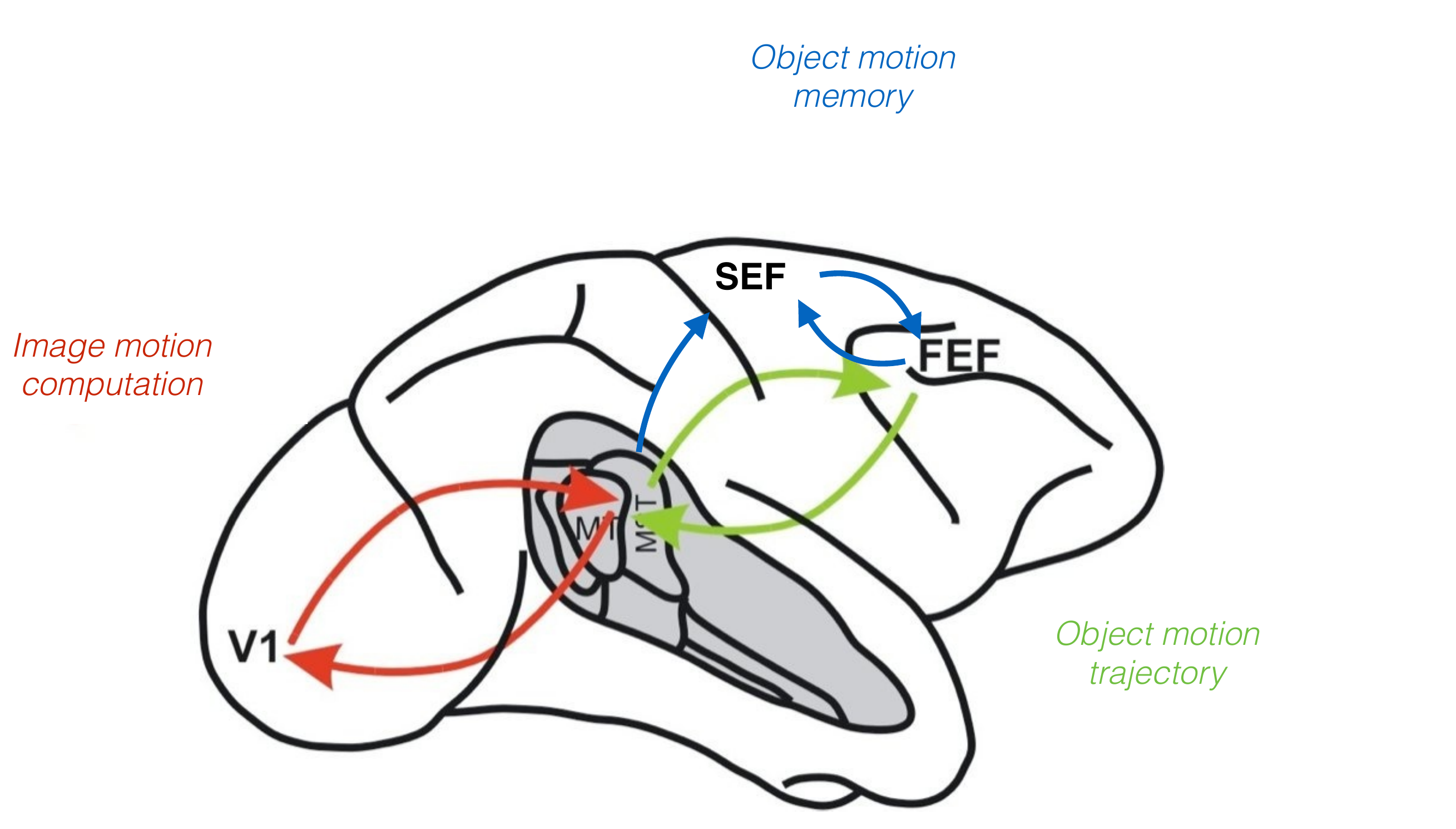}
}
\caption{A lateral view of the macaque cortex. The neural network corresponding to our hierarchical Bayesian model of smooth pursuit is made of three main cortico-cortical loop. The first loop between primary visual cortex (V1) and the medio-temporal (MT) computes image motion and infers the optimal low-level solution for object motion direction and speed. Its main output is the medio-superior temporal (MST) area that acts as a gear between the sensory loop and the object motion computation loop. Retinal and extra-retinal signals are integrated in both MST and FEF areas. Such dynamical integration computes the perceived trajectory of the moving object and implements an online prediction that can be used on the course of a tracking eye movement to compensate for target perturbation such as transient blanking. FEF and MST areas signals are sent to the supplementary eye field (SEF) and the interconnected prefrontal cortical areas. This third loop can elaborate a motion memory of the target trajectory and is interfaced with higher cognitive processes such as cue instruction or reinforcement learning. It also implements offline predictions that can be used across trials, in particular to drive anticipatory responses to highly predictable targets.
\label{fig:neural_workspace}}% 
\end{figure}%

In non-human primates, the medial superior temporal (MST) cortical area is essential for pursuit. It receives MT inputs about target direction and speed of visual pattern and represents the target's velocity vector. In the context of smooth pursuit control, neuronal activities in area MST show several interesting features. First, during pursuit, if the target is blanked, the neuronal activity is maintained throughout the course of the blanking. This is clearly different from area MT where neurons stop firing in the occurrence of even a brief disappearance of the target~\citep{Newsomeetal1988}. Second, both monkeys and humans can track imaginary large line-drawing targets where the central, foveal part is absent~\citep{IlgThier1999}. MST neurons, but not MT cells, can signal the motion direction of these parafoveal targets, despite the fact visual edges fall outside their receptive fields. Thus, in area MST, neurons are found whose activities are not different during pursuit of real (i.e. complete) or imaginary (i.e. parafoveal) targets~\citep{Ilg03,MassonIlg2010}. Lastly, many MST neurons can encode target motion veridically during eye movements in contrast to MT cells~\citep{ChukoskieMovshon:2009}. All these evidences strongly suggest that MST neurons integrate both retinal and extra-retinal information to reconstruct the perceived motion of the target. However, despite the fact that areas MT and MST are strongly, and recurrently connected, the strong difference between MT and MST neuronal responses between pursuit seems to indicate that extra-retinal signals are not back propagated to early visual processing areas. Interestingly, several recent models have articulated their two-stage computational approach with this architecture~\citep{pack:01, grossberg:12} in order to model the dynamics of primate smooth-pursuit.

From there, output signals are sent in two directions. One signal reaches the brainstem oculomotor structures through the pontine nuclei, the cerebellar floccular region and the vestibular nuclei~\citep{LeighZee:06}. This cortico-subcortical pathway conveys the visual drive needed for pursuit initiation and maintenance. The second signal reaches the frontal cortex that includes the caudal parts of the frontal eye fields (FEF) and the supplementary eye field (SEF) (see~\citep{Krauzlis2004,Fukushima13} for recent reviews). FEF neurons share many properties of MST cells. In particular, they integrate both retinal and extra-retinal information during pursuit so that responses remain sustained during blanking or when simulated with imaginary targets~\citep{Ilg03,MassonIlg2010b}. Moreover, both FEF and MST cells show a build-up of activity during anticipatory pursuit~\citep{Ilg2003,Fukushima02}. Thus, FEF and MST appear to be strongly coupled to build an internal representation of target motion that can be used during steady-state tracking as well as during early phases of pursuit. Moreover, FEF area issues pursuit commands that are sent to the brainstem nucleus reticularis tegmenti pontis (NRTP) and the cerebellar vermis lobules before reaching the pursuit oculomotor structures. Several authors have proposed that such parieto-frontal loop might implement the prediction component of the pursuit responses  (see~\citep{MassonIlg2010b} for review). Other have proposed to restrict its role to the computation of the perceived motion signal that drive the pursuit response~\citep{Fukushima13} while higher signals related to prediction might be computed in more anterior areas such as SEF and prefrontal cortex (PFC). 

Prediction is influenced by many cognitive inputs (cues, working memory, target selection)~\citep{Barnes2008}. Accordingly, prediction-related neuronal responses during pursuit have been reported in area SEF~\citep{Heinen95} and the caudal part of FEF~\citep{Fukushima02}. Moreover, SEF activity facilitates anticipatory pursuit responses to highly predictable targets~\citep{MissalHeinen:04}. The group of Fukushima have identified several subpopulations of neurons in both areas that can encode directional visual motion-memory, independently of movement preparation signals (see~\citep{Fukushima13} for a complete review). However, FEF neurons are more often mixing predictive and motor preparation signals while SEF cells more specifically encode a visual motion memory signal. This is consistent with the fact that many neurons in the prefrontal cortex have been linked to temporal storage of sensory signals~\citep{GoldmanRakic:95}. Thus, a working memory of target motion might be formed in area SEF by integrating multiple inputs from parietal (MST) and prefrontal (FEF, PFC) cortical areas. \Citet{Fukushima13} proposed that a recurrent network made of these areas (MST, FEF, SEF, PFC) might signal future target motion using prediction, timing, expectation as well as experience gained over trials.

All these studies define a neuronal workspace for our hierarchical inference model, as illustrated in Figure~\ref{fig:neural_workspace}. Two main loops seem to be at work. A first loop predicts the optimal target motion direction and speed from sensory evidence (image motion computation, in red). It uses sensory priors such as the "smooth and slow motion prior" used for both perception and pursuit initiation that are largely immune to higher influence. By doing so, the sensory loop can preserve its ability to quickly react to new sensory event and avoid the inertia of prediction systems. This loop would correspond to the reactive pathway of the pursuit model proposed by \citet{Barnes2008,Fukushima13}. On the ground of some  behavioral evidence \citep{Montagnini06}, we believe that target motion prediction can not easily overcome the aperture problem (see also the open questions discussed in section \ref{sec:open_questions}), providing a strong indication that this sensory loop is largely not penetrable to cognitive influences. The second loop involves areas MST and FEF to compute and store online target motion by taking into account both sensory and efference copy signals (object motion computation, green). A key aspect is that MST must act as a gear to prevent predictive or memory-related signals to back propagate downstream to the sensory loop. We propose to distinguish between online prediction, involving these two areas during on-going event such as target blanking to offline prediction. The later is based on a memory of target motion that span across trials and might be used to both trigger anticipatory responses or drive responses based on cues. It might most certainly involve a dense, recurrent prefrontal network articulated around SEF and that offers a critical interface with the cognitive processes interfering with pursuit control (Object memory loop, in blue).

This architecture presents many advantages. First, it preserves the brain to quickly react to new event as a brutal change in target motion direction. Second, it ensures maintaining pursuit in a wide variety of conditions with good stability and by constructing an internal model of target motion it allows a tight coordination between pursuit and saccades~\citep{OrbanLefevre:07}. Third, it provides an interface with higher cognitive aspects of sensorimotor transformation. Several questions remain however unsolved. Because of the strong changes seen in prediction with different behavioral context, several models, including the one presented here, postulate the existence of hard switches that can turn on or off the contribution of a particular model component. We need a better theoretical approach about decision making between these different loops. The Bayesian approach proposed here, similar to the Kalman filter models, opens the door to better understanding these transitions. It proposes that each signal (sensory evidence, motion memory, prediction) can be weighted from its reliability. Such unifying theoretical approach can then be used to design new behavioral and physiological experiments.

%-----------------------------------------------------------------------------------------------------------------------------------%
\section{Conclusion}
\label{sec:open_questions}

Recent experimental evidence points to the need to revise our view of the primates' smooth pursuit system. Rather than a reflexive velocity-matching negative-feedback loop, the human motion tracking system seems to be grounded on a complex set of dynamic functions that subserve a quick and accurate adaptive behavior even in visually challenging situations. By analyzing tracking eye movements produced with a simple, unique and highly-visible moving target, many of these notions could not have clearly emerged and it is now clear that testing more naturalistic visual motion contexts and carefully taking into account the sources of uncertainty at different scales is a crucial step for understanding biological motion tracking.  The approach highlighted here opens the door to several open questions.

First, we have focused herein on luminance-based motion processing mechanisms. Such inputs can be well extracted by a bank of filters extracting motion energy at multiple scales. The human visual motion system is however more versatile and psychophysical studies have demonstrated that motion perception is based on cues that can be defined in many different ways. These empirical studies have led to the three-systems theory of human visual motion perception by~\citet{LuSperling01}. Beside the first-order system that responds to moving luminance patterns, a second-order system responds to moving modulations of feature types (i.e. stimuli where the luminance is the same everywhere but an area of higher contrast or of flicker moves). A third-order system slowly computes the motion of marked locations in a ``salience map'' where locations of important visual features in the visual space (i.e. a figure) are highlighted respective to its ``background'' (see Chapter \verb+013_lemeur+). The contribution of the first-order motion to the initiation of tracking responses have been largely investigated (see~\citep{Masson12, Lisberger2010} for reviews). More recent studies have pointed out that feature tracking mechanisms (i.e. a second-order system) are critical for finely adjusting this initial eye acceleration to object speed when reaching steady-state tracking velocity~\citep{wilmer07}. Whether the third-order motion system is involved, and how in the attentional modulation of tracking, is currently under investigation by many research groups. Altogether, these psychophysical and theoretical studies point towards the need for more complex front-end layers of visuomotor models so that artificial systems would more versatile and adapted to complex, ambiguous environments.

Second, although the need of hierarchical, multi-scale inferential models is now apparent, current models will have to meet the challenge of explaining a rich and complex set of behavioral data. Just to detail an example, somewhat unexpectedly we have found that predicting 2D motion trajectories does not help solving the aperture problem~\citep{Montagnini06}. Indeed, when human subjects are exposed to repeated motion conditions for a horizontally-translating tilted bar across several experimental trials and they develop anticipatory smooth pursuit in the expected direction, still their pursuit traces reflect the aperture-induced bias at initiation, as illustrated in Figure~\ref{fig:aperture_SP}, panel $d$. It is important to notice that the robust optimal cue combination  of the output of sensory and predictive motion processing modules postulated in the previous section~\citep{Orban2013,Bogadhi13} are not capable, at this stage, to explain this phenomenon. Other crucial issues deserve to be better understood and modeled in this new framework, such as, for instance, the dynamic interaction between global motion estimate and object segmentation, the precise functional and behavioral relationship between smooth tracking and discrete jump-like saccadic movements, or the role of high-level cognitive cues in modulating motion tracking~\citep{Kowler14}.

\subsection{Interest for computer vision}
Machine-based motion tracking, similar to human motion tracking, needs to be robust to visual perturbations. As we have outlined above, including some form of predictive information might be extremely helpful to stabilize object tracking, in much the same way as what happens during transient target blanking for human smooth pursuit. Importantly, human tracking seems to be equipped with a more complete and advanced software for motion tracking, namely one that allows us i) to exploit the learned regularities of motion trajectories to anticipate and compensate for sensorimotor delays and consequent mismatch between gaze and target position in the event of a complex trajectory; ii) to keep tracking an object temporarily occluded by other objects; and iii) to start tracking in the dark when the future motion of an object is predictable, or even partly predictable. Some of these predictive cues might depend upon a high-level cognitive representation of kinematic rules or arbitrary stimulus-response associations. This type of apparatus is probably much too advanced to be implemented into a common automatic motion tracking device, but still it may be a source of inspiration for future advanced developments for machine vision and tracking. 
%

%------------------%

\subsection*{Acknowledgments}  %
\Acknowledgments %
\bibliography{biblio_motiontracking}%
%\include{c02}
%...
%\appendix
%\include{a01}
\backmatter
%\include{b01}
%\printindex
\end{document}